\documentclass[apj,a4paper,12pt,useAMS]{emulateapj} 

\usepackage{amsmath, amssymb, amsfonts}

\shorttitle{AMiBA: Scaling Relations}
\shortauthors{Huang et al.}

\begin{document}

\title{AMiBA: SCALING RELATIONS BETWEEN THE INTEGRATED COMPTON-$y$ AND X-RAY
DERIVED TEMPERATURE, MASS, AND LUMINOSITY}

\author{Chih-Wei Locutus Huang\altaffilmark{1,2}, Jiun-Huei Proty Wu\altaffilmark{1,2},
Paul T. P. Ho\altaffilmark{3,4}, Patrick M. Koch\altaffilmark{3},
Yu-Wei Liao\altaffilmark{1,2}, Kai-Yang Lin\altaffilmark{1,3},
Guo-Chin Liu\altaffilmark{3,5}, Sandor M. Molnar\altaffilmark{3},
Hiroaki Nishioka\altaffilmark{3}, Keiichi Umetsu\altaffilmark{2,3},
Fu-Cheng Wang\altaffilmark{1,2}, Pablo Altamirano\altaffilmark{3},
Mark Birkinshaw\altaffilmark{6}, Chia-Hao Chang\altaffilmark{3},
Shu-Hao Chang\altaffilmark{3}, Su-Wei Chang\altaffilmark{3},
Ming-Tang Chen\altaffilmark{3}, Tzihong Chiueh\altaffilmark{1,2},
Chih-Chiang Han\altaffilmark{3}, Yau-De Huang\altaffilmark{3},
Yuh-Jing Hwang\altaffilmark{3}, Homin Jiang\altaffilmark{3},
Michael Kesteven\altaffilmark{7}, Derek Kubo\altaffilmark{3},
Chao-Te Li\altaffilmark{3}, Pierre Martin-Cocher\altaffilmark{3},
Peter Oshiro\altaffilmark{3}, Philippe Raffin\altaffilmark{3},
Tashun Wei\altaffilmark{3}, Warwick Wilson\altaffilmark{7}}

\altaffiltext{1}{Department of Physics, Institute of Astrophysics,
\& Center for Theoretical Sciences, National Taiwan University, Taipei
10617, Taiwan} \altaffiltext{2}{LeCosPA Center, National Taiwan
University, Taipei 10617, Taiwan} \altaffiltext{3}{Institute
of Astronomy and Astrophysics, Academia Sinica, P.~O.~Box 23-141,
Taipei 10617, Taiwan} \altaffiltext{4}{Harvard-Smithsonian Center
for Astrophysics, 60 Garden Street, Cambridge, MA 02138, USA} \altaffiltext{5}{Department
of Physics, Tamkang University, 251-37 Tamsui, Taipei County, Taiwan}
\altaffiltext{6}{Department of Physics, University of Bristol,
Tyndall Avenue, Bristol BS8 1TL, UK} \altaffiltext{7}{Australia
Telescope National Facility, P.~O.~Box 76, Epping NSW 1710, Australia}

\begin{abstract}
We investigate the scaling relations between the X-ray and the thermal
Sunyaev--Zel'dovich Effect (SZE) properties of clusters of galaxies,
using data taken during 2007 by the Y.T. Lee Array for Microwave Background
Anisotropy (AMiBA) at 94~GHz for the six clusters A1689, A1995, A2142,
A2163, A2261, and A2390. The scaling relations relate the integrated
Compton-$y$ parameter $Y_{2500}$ to the X-ray derived gas temperature $T_\textrm{e}$, 
total mass $M_{2500}$, and bolometric luminosity $L_X$ within $r_{2500}$. 
Our results for the power-law index and normalization are both consistent with
the self-similar model and other studies in the literature except for the $Y_{2500}$--$L_X$ relation,
for which a physical explanation is given though further investigation may be still needed.
Our results not only provide confidence for the AMiBA project
but also support our understanding of galaxy clusters.
\end{abstract}


\section{INTRODUCTION}

The Sunyaev--Zel'dovich Effect (SZE) is a powerful tool that can potentially
answer long-standing questions about the large-scale distribution
of matter. The SZE is a spectral distortion of the cosmic microwave
background (CMB), induced when a fraction of CMB photons are scattered
by hot electrons in the cores of massive galaxy clusters \citep{SZ1970}.
The redshift independence of the SZE  enables the direct detection of distant
clusters without the $(1+z)^{4}$ surface brightness dimming that limits other
techniques, including X-ray observations. Clusters studied via the
SZE are therefore effective cosmological probes. Studying their properties
in detail will lead to heightened understanding of the mass power
spectrum, and should provide improved constraints on cosmological
parameters.

In the simplest scenario, where gravity is assumed to be the only
influence on the formation of galaxy clusters, a simple `self-similar'
model can be used to relate the physical properties of clusters \citep{Kaiser1986}.
Assuming spherical collapse of the dark matter (DM) halo, and hydrostatic
equilibrium of gas in the DM gravitational potential, one can derive
power-law scaling relations between various X-ray and SZE quantities,
e.g.~luminosity and temperature, gas mass and temperature, total
mass and luminosity, entropy and temperature, and Compton-$y$ parameter
and temperature. Existence of these relations in observations can
be seen in, for example, \citet{Mushotzky1997,Benson2004,Bonamente2007,Morandi2007}.
These relations are also found in numerical simulations, e.g., between
X-ray quantities \citep{Nagai2007}, SZE flux and total mass or gas
mass \citep{Motl2005,Nagai2006}, and integrated Compton-$y$ and
temperature or luminosity \citep{Silva2004}. Deviations from the
scaling relations should reveal the importance of non-gravitational
processes for the formation of clusters \citep[e.~g.~][]{Allen1998a,McCarthy2002,McCarthy2003a},
or constrain the mass distributions of clusters \citep{Reiprich2002}.
Furthermore, the scaling relations can be used as an utility to extract
important quantities and evolution behaviors for remote clusters using
SZE observables alone.

The Y.T.~Lee Array for Microwave Background Anisotropy (AMiBA) experiment
\citep{Ho2008} observed and detected the SZEs of six massive Abell
clusters in the range $0.09<z<0.32$ during the year 2007 \citep{Wu2008}.
AMiBA is a coplanar interferometer that during 2007 operated at 94~GHz
with seven 0.6-m antennas in a hexagonal close-packed configuration,
giving a synthesized resolution of about $6\arcmin$.
The array has a sensitivity of 63~mJy/hr for on-source integration, 
and an overall efficiency of $0.36$ \citep{Lin2008}. 
Details for the transformation of the raw data into calibrated
visibilities are presented in \citet{Wu2008}, and the checks on data
integrity are described in \citet{Nishioka2008}. At our observing
frequency the SZE signal is an intensity decrement in the CMB. We
fit the central (peak) decrement in the AMiBA visibilities using isothermal
$\beta$-models \citep{CF1976}, taking account of contamination from
the primary CMB and foreground emissions \citep{Liu2008}. Other companion
papers include \citet{Chen2008} and \citet{Koch2008a}, where the
technical aspects of the instruments are described, \citet{Umetsu2008},
where the AMiBA SZE data is combined with weak lensing data from Subaru
to analyze the distributions of mass and hot baryons, \citet{Koch2008},
where the Hubble constant is estimated from AMiBA SZE and X-ray data,
and \citet{Molnar2008}, which discusses the feasibility of further constraining
the intra-cluster gas model using AMiBA upgraded to 13 antennas \citep[AMiBA13]{Ho2008}.
The consistency of our results with other observations and theoretical
expectations will validate not only the performance and capability
of instruments, but also the analysis methodology. Since AMiBA is
one of few leading SZE instruments operating at 3-mm wavelength, we
anticipate that it will fill an important role by providing 3-mm SZE
data for spectral studies.

In this article we address the scaling relations between the integrated SZE Compton-$y$
parameter obtained by AMiBA and X-ray gas temperature, X-ray luminosity,
and total cluster mass derived from the literature. 
In Section~\ref{sec:xrpars}
we discuss the cluster gas models and cluster parameters derived from the X-ray data.
In Section~\ref{sec:szepars} 
we calculate the integrated Compton-$y$ parameter for each of the six clusters. 
In Section~\ref{sec:scalings}
we investigate the scaling relations including the consideration for errors. 
We further discuss our results in Section~\ref{sec:dandc} and
draw conclusions in Section~\ref{sec:conclusions}.

\section{CLUSTER PROPERTIES FROM X-RAY DATA}
\label{sec:xrpars}
As the $u$--$v$ coverage is incomplete for a single interferometric
SZE experiment, we can not measure the accurate profile of a cluster
or its central intensity. Therefore we have chosen to assume a cluster
model, and thus a flux-density profile, so that a corresponding template
in $u$--$v$ space can be fitted to the observed visibilities in
order to estimate the underlying model parameters including the
central SZE intensity, $\Delta I_{0}$. We apply the spherical isothermal
$\beta$-model in our X-ray and SZE analysis. The cluster gas density
distribution is of the form 
\begin{equation}
	n_{\mathrm{e}}(r)=n_{\mathrm{e}0}\left(1+\frac{r^{2}}{r_{\mathrm{c}}^{2}}\right)^{-3\beta/2}\quad,
\end{equation}
where $n_{\mathrm{e}0}$ is the central number density of electrons,
$r$ is the radius from the cluster center, $r_{\mathrm{c}}$ the
core radius and $\beta$ is a structure index. Due to the limited resolution of AMiBA in its 7-element closed-packed configuration, we cannot obtain good estimates for some of the model parameters from our SZE data alone.
Therefore we have taken the X-ray derived values for $\beta$ and $\theta_{\mathrm{c}}$ from the literature,
where $\theta_{\mathrm{c}}=r_{\mathrm{c}}/D_{\mathrm{A}}$ and $D_{\mathrm{A}}$ is the angular diameter distance.
Throughout this paper,
we assume a flat $\Lambda$CDM universe
with $H_{0}=70$ km $\mathrm{s}^{-1}\mathrm{Mpc}^{-1}$, $\Omega_{\mathrm{M}}=0.3$, and $\Omega_{\Lambda}=0.7$.

To relate with the SZE Compton-$y$ parameter, we also need to borrow the cluster gas temperature $T_{\mathrm{e}}$, the total mass $M_{2500}$, and the bolometric luminosity $L_{\mathrm{X}}$ derived from the X-ray data. $M_{2500}$ refers to the total mass in a cluster central region out to $r=r_{2500}$, defined
as the radius of mean overdensity $2500\times\rho_{\mathrm{c}}$ where
$\rho_{\mathrm{c}}$ is the critical density at redshift $z$. Given $\beta$, $r_{\mathrm{c}}$ and $T_{\mathrm{e}}$ for a cluster, we can compute first $r_{2500}$ and then $M_{2500}$ through the total mass
equation of the $\beta$-model \citep{Grego2001}
\begin{equation}
	M_{2500}=2500\frac{4\pi}{3}r_{2500}^{3}\rho_{\mathrm{c}}=\frac{3\beta k_\textrm{B}T_{\mathrm{e}}}{G\mu 			m_{p}}\frac{r_{2500}^{3}}{r_{\mathrm{c}}^{2}+r_{2500}^{2}}\quad,\label{eq:total mass}
\end{equation}
where $\mu=0.6$ is the mean molecular weight in units of $m_{\mathrm{H}}$ for an assumed near-solar metallicity of the intracluster medium.

We considered two sets of X-ray derived parameters.
The first set is mainly based on the \textit{Chandra} data, 
and this leads to our main results.
The second set is mainly derived from \textit{ROSAT}
images or a combination of \textit{ROSAT} data and \textit{ASCA}
spectral measurements (the \textit{ASCA/ROSAT} parameters in what
follows). Because these data are generally of lower accuracy,
we include them only for comparison.

\subsection{Chandra}
To deal with the complicated non-gravitational physics in cluster
cores, including radiative cooling and feedback mechanisms, and
the transient boosting of surface brightness and spectral temperature
during merging events, the parameters of the \textit{Chandra} set were
derived by fitting an isothermal $\beta$-model to the X-ray data with
the central 100 kpc excised. The most recent and currently most
extensive studies of $H_{0}$ \citep{Bonamente2006} and the gas mass
fraction $f_{\mathrm{gas}}$ \citep{LaRoque2006} adopted this 100-kpc
cut model in their analysis, and claim that a cut at 100 kpc is 
large enough to exclude the cooling region in cool-core clusters while
retaining sufficient photons for modeling. This model was also used in
recent studies of scaling relations based on X-ray and SZE
observations \citep{Morandi2007,Bonamente2007}.

Table~\ref{tab:X-obs-C} summarizes
the parameters from \textit{Chandra} observations, and the
values of $r_{2500}$ and $M_{2500}$ derived from them. 
Note that \textit{Chandra}-based values of
$\beta$ and $\theta_{\mathrm{c}}$ for A2142 are unavailable in the literature,
and so we adopted values taken from the \textit{ASCA/ROSAT} set, which
were not determined by fitting the 100-kpc cut model. The gas
temperature for A2142 is \textit{Chandra}-based, as given by
\citet{Markevitch2000}. The temperature fit allowed a cooling
component to be present, but was based on the overall X-ray spectrum
of A2142, rather than discarding photons extracted from the
central 100 kpc region. Nevertheless, if A2142 is excluded from the
sample for this set of parameters, it has only a minor effect on the
scaling relations (less than a $5\%$ change for the power index or the
normalization; see Sec.~\ref{sec:scalings}).

The values of $\beta$ and $\theta_{\mathrm{c}}$ for A2390 are taken
from \citet{Allen2001} who fit the X-ray surface brightness profile to
an isothermal $\beta$-model between radii $80$~and $900$~kpc.
As the authors remark, however, an isothermal $\beta$-model
ignoring the central region associated with the possible cooling flow 
cannot describe the mass distribution well, since there is a
`break' in the surface brightness profile at $r\sim500$
kpc. A better fit can be obtained using a simple broken power-law
model, or assuming an NFW \citep{Navarro1997} potential with the
assumption of gas isothermality relaxed.

\subsection{ASCA/ROSAT}
Parameters derived from \textit{ASCA} and \textit{ROSAT} are summarized in
Table~\ref{tab:X-obs-R}. The gas temperatures and the bolometric
luminosities of our clusters, except A1995 and A2163, are
compiled by \citet{Allen1998a} and \citet{Allen2000}, where the X-ray
spectra were fitted by using a model with an isothermal plasma in
collisional equilibrium, including an additional component explicitly
to account for cooling flows (Model~C). For A2163, which is
not a cooling-core cluster, we take the values from the same papers,
but without the additional cooling component (Model~A). For
A1995, which is absent from these papers, we use the value
of $T_{\mathrm{e}}$ from \citet{Patel2000}, who detected no
excess in the X-ray surface brightness biased from a cooling flow in
the cluster center. However, A1995 has recently been classified as a
cooling cluster, according to the criterion that the cooling time in
the central inner region is less than the Hubble time at the cluster
redshift \citep{Morandi2007}.

All values in Tables~\ref{tab:X-obs-C} and \ref{tab:X-obs-R} are presented at the $68.3\%$ confidence level.
Errors are obtained by propagating the errors in the input parameters from the literature
through a Monte-Carlo process.

\begin{deluxetable*}{l|c|ccccccc}
\tabletypesize{\scriptsize}
\tablewidth{0pt}
\tablecaption{Cluster parameters of the \textit{Chandra} set
\label{tab:X-obs-C}} 
\tablehead{
        & $z$\tablenotemark{a} & $\beta$ & $\theta_{\mathrm{c}}$ 
& $r_{2500}$ & $T_{\mathrm{e}}$      & $M_{2500}$          & $L_{\mathrm{X}}$ & ref\\
Cluster &     &         & (\arcsec)
& (kpc)      & (keV)\tablenotemark{c}& ($10^{14}M_{\sun}$) & ($10^{45}$ erg/s)\tablenotemark{c} & ($\beta$ \& $\theta_{\mathrm{c}}$, $T_{\mathrm{e}}$, $L_{\mathrm{X}}$)
}
\startdata
A1689  & $0.183$ 
       & $0.686_{-0.01}^{+0.01}$ & $48.0_{-1.7}^{+1.5}$ & $607_{-23}^{+22}$ & $8.72_{-0.56}^{+0.63}$ & $3.82_{-0.42}^{+0.43}$ & $3.15\pm0.09$ & 1, 4, 4\\
A1995  & $0.322$ 
       & $0.923_{-0.023}^{+0.021}$ & $50.4_{-1.5}^{+1.4}$ & $579_{-21}^{+21}$ & $7.56_{-0.41}^{+0.45}$ & $3.87_{-0.41}^{+0.43}$ & $1.51\pm0.05$ & 1, 4, 4\\
A2142\tablenotemark{b}  & $0.089$ 
       & $0.74_{-0.01}^{+0.01}$ & $188.4_{-13.2}^{+13.2}$ & $608_{-31}^{+30}$ & $8.80_{-0.55}^{+0.73}$ & $3.49_{-0.52}^{+0.53}$ & -- & 2, 5, --\\
A2163  & $0.202$ 
       & $0.700_{-0.07}^{+0.07}$ & $78.8_{-0.6}^{+0.6}$ & $684_{-41}^{+40}$ & $12.0_{-0.26}^{+0.28}$ & $5.59_{-0.96}^{+1.00}$ & $4.80\pm0.05$ & 1, 4, 4\\
A2261  & $0.224$ 
       & $0.628_{-0.02}^{+0.03}$ & $29.2_{-2.9}^{+4.8}$ & $531_{-22}^{+22}$ & $7.47_{-0.47}^{+0.53}$ & $2.67_{-0.32}^{+0.33}$ & $2.02\pm0.07$ & 1, 4, 4\\
A2390  & $0.233$ 
       & $0.58_{-0.058}^{+0.058}$ & $43.3_{-4.33}^{+4.33}$ & $583_{-33}^{+32}$ & $10.18_{-0.21}^{+0.23}$ & $3.57_{-0.58}^{+0.61}$ & $4.66\pm0.05$ & 3, 4, 4
\enddata
\tablerefs{(1) \citet{Bonamente2006}. (2) \citet{Sanderson2003,Lancaster2005}. (3) \citet{Allen2001} with a $10\%$ error assumed. (4) \citet{Morandi2007}. (5) \citet{Markevitch2000}.}
\tablenotetext{a}{The redshifts $z$ are from \citet{Bonamente2006} except those for A2142 \& A2390 which are given by \citet{Allen2000}.}
\tablenotetext{b}{We take the values of $\beta$ and $\theta_{\mathrm{c}}$ used in the \textit{ASCA/ROSAT} set (Table~\ref{tab:X-obs-R}) for A2142 since they are not available in
 the \textit{Chandra}-based literature.}
\tablenotetext{c}{The emission-weighted temperatures and the bolometric luminosities are extracted in a region of radius $r$ between 100 kpc and $r_{2500}$.}
\end{deluxetable*}

\begin{deluxetable*}{l|c|ccccccc}
\tabletypesize{\scriptsize}
\tablewidth{0pt}
\tablecaption{Cluster parameters of the \textit{ASCA/ROSAT} set
\label{tab:X-obs-R}} 
\tablehead{
        & $z$ & $\beta$ & $\theta_{\mathrm{c}}$ 
& $r_{2500}$ & $T_{\mathrm{e}}$ & $M_{2500}$ & $L_{\mathrm{X}}$ & ref\\
Cluster &     &         & (\arcsec)
& (kpc)      & (keV)            & ($10^{14}M_{\sun}$) & ($10^{45}$ erg/s) & ($\beta$ \& $\theta_{\mathrm{c}}$, $T_{\mathrm{e}}$, $L_{\mathrm{X}}$)
}
\startdata
A1689  & $0.183$ 
       & $0.609_{-0.005}^{+0.005}$ & $26.6_{-0.7}^{+0.7}$ & $625_{-20}^{+19}$ & $10.0_{-0.49}^{+0.73}$  & $4.17_{-0.38}^{+0.39}$  & $6.26$ & 1, 4, 4\\
A1995  & $0.322$ 
       & $0.770_{-0.063}^{+0.117}$ & $38.9_{-4.3}^{+6.9}$ & $579_{-48}^{+47}$ & $8.59_{-0.67}^{+0.86}$  & $3.88_{-0.91}^{+1.00}$  & -- & 1, 5, --\\
A2142  & $0.089$ 
       & $0.74_{-0.01}^{+0.01}$ & $188.4_{-13.2}^{+13.2}$ & $629_{-29}^{+28}$ & $9.3_{-0.43}^{+0.79}$  & $3.87_{-0.51}^{+0.53}$  & $6.78$ & 2, 4, 4\\
A2163  & $0.202$ 
       & $0.674_{-0.008}^{+0.011}$ & $87.5_{-2.0}^{+2.5}$ & $716_{-15}^{+15}$ & $13.83_{-0.45}^{+0.47}$  & $6.37_{-0.41}^{+0.42}$ & $14.7$ & 1, 4, 4\\
A2261  & $0.224$ 
       & $0.516_{-0.013}^{+0.014}$ & $15.7_{-1.1}^{+1.2}$ & $589_{-71}^{+66}$ & $10.9_{-1.34}^{+3.59}$  & $3.68_{-1.20}^{+1.31}$ & $5.83$ & 1, 4, 4\\
A2390  & $0.233$ 
       & $0.6_{-0.06}^{+0.06}$ & $28_{-2.8}^{+2.8}$ & $721_{-174}^{+155}$ & $14.5_{-3.16}^{+9.42}$  & $7.09_{-4.14}^{+5.02}$ & $10.11$ & 3, 4, 4
\enddata
\tablerefs{(1) \citet{Reese2002}. (2) \citet{Sanderson2003,Lancaster2005}. (3) \citet{Bohringer1998} with a $10\%$ error assumed. (4) \citet{Allen2000}. (5) \citet{Patel2000}.}

\end{deluxetable*}

\section{CLUSTER PROPERTIES FROM SZE}
\label{sec:szepars}
In AMiBA targeted observations at 94~GHz, the sky signal is dominated
by the thermal SZE. 
The amplitude of such signals is proportional to the Compton-$y$ parameter,
$y=\sigma_{T}/\left(m_{\mathrm{e}}c^{2}\right)\int_{0}^{\infty}k_\textrm{B}T_{\mathrm{e}}(l)n_{\mathrm{e}}(l)\: dl$
where $\sigma_{T}$ is the Thomson scattering cross section, $k_\textrm{B}T_{\mathrm{e}}(l)n_{\mathrm{e}}(l)$
is the electron pressure, and the integral is taken along the line
of sight. The Compton-$y$ parameter can be interpreted as a measure
of Comptonization integrated through a cluster. In terms of a change
in intensity, the thermal SZE observed at frequency $\nu$ can be represented by a decrement
\begin{equation}
	\Delta I_\textrm{SZE}=y\cdot g\left(x,T_{\mathrm{e}}\right)\cdot I_\textrm{CMB}\,,\label{eq:sz-intensity}
\end{equation}
where $x\equiv h\nu/\left(k_\textrm{B}T_\textrm{CMB}\right)$,
$T_\textrm{CMB}=2.725$~K \citep{Mather1999},
and 
$I_\textrm{CMB}\equiv2h\nu^{3}\mathrm{c}^{-2}\left(e^{x}-1\right)^{-1}$
is the CMB intensity. 
The factor $g\left(x,T_{\mathrm{e}}\right)$ can be expressed as \citep{Bonamente2007,Morandi2007,Udomprasert2004}
\begin{equation}
	g\left(x,T_{\mathrm{e}}\right)=\frac{xe^{x}}{e^{x}-1}\left(F-4\right)
	+\delta_\textrm{rel}\left(x,T_{\mathrm{e}}\right),
\label{eq:freqfactor}
\end{equation}
where $\delta_\textrm{rel}\left(x,T_{\mathrm{e}}\right)$
is a small relativistic correction \citep{Challinor1998}
\begin{equation}
	\label{eq:relativistic}
	\begin{split}
	\delta_\textrm{rel}\left(x,T_{\mathrm{e}}\right) &= 
	\frac{xe^{x}}{e^{x}-1}
	\frac{k_\textrm{B}T_{\mathrm{e}}}{m_{\mathrm{e}}c^{2}}
	\Bigl[-10+\frac{47}{2}F-\frac{42}{5}F^{2}+\\
	& \quad \frac{7}{10}F^{3}+\frac{7}{5}G^{2}(-3+F)\Bigr],
	\end{split}
\end{equation}
$F\equiv x\coth(x/2)$, and $G\equiv x/\sinh(x/2)$. 
The relativistic correction is about $6\%$ for $\nu=94$~GHz and $T_{\mathrm{e}}=10$~keV,
which is a typical temperature for our SZE clusters.

Given a gas density profile $n_{\mathrm{e}}(r)$ we can determine
the distribution of $\Delta I_\textrm{SZE}$ on the plane of the sky. For an isothermal $\beta$-model,
the projected SZE decrement distribution has a simple analytical form
\citep[e.g.][]{Udomprasert2004}
\begin{equation}
	\Delta I_{{\rm SZE}}(\theta)=\Delta I_{0}\left(1+\frac{\theta^{2}}{\theta_{\mathrm{c}}^{2}}\right)^{(1-3\beta)/2},
\end{equation}
where $\theta$ and $\theta_{\mathrm{c}}$
are the angular equivalents of $r$ and $r_{\mathrm{c}}$ respectively, 
and $\Delta I_{0}$ is the central SZE intensity decrement. Because
the SZE clusters are not well resolved by AMiBA, we cannot get
a good estimate of $\Delta I_{0}$, $\beta$, and $\theta_{\mathrm{c}}$
simultaneously from our data alone. Instead, we adopt the X-ray derived
values for $\beta$ and $\theta_{\mathrm{c}}$ from \textit{Chandra} or \textit{ASCA/ROSAT}, 
and then estimate $\Delta I_{0}$ \citep{Liu2008} by fitting the
$\beta$-model to the SZE visibilities obtained in \citet{Wu2008}.

For the two different sets of X-ray parameters we accordingly obtain
two sets of $\Delta I_{0}$ values. For the the \textit{Chandra} set
with a 100 kpc-cut model we choose to fit the entire SZE data, while
using the X-ray parameters from the same model. \citet{LaRoque2006}
and \citet{Bonamente2006} already remarked that there is no simple way to
mask the central 100 kpc from the interferometric SZE data because these data
are in the $u$--$v$ space. 
Nevertheless, our approach should be valid because the limited resolution of
the current AMiBA is insensitive to the details of the cluster core.
Moreover, since the SZE probes the integrated gas pressure,
which is linear in $n_{\mathrm{e}}$,
the parameters derived from the SZE data should be less dependent on the
core properties than parameters derived from X-ray observations,
where the X-ray surface brightness $\propto n_{\mathrm{e}}^2$.
Table~\ref{tab:SZ-obs} summarizes
the resulting estimated values of $\Delta I_{0}$ based on \textit{Chandra}.
We note that
the effects of foregrounds such as radio source contamination, Galactic emission, and
confusion from primary CMB fluctuations have been taken into account \citep{Liu2008}.

In addition to the intensity decrement, the thermal SZE also can be
expressed in terms of a change of the thermodynamic temperature of the CMB, 
$\Delta T_\textrm{SZE}=y \cdot
g\left(x,T_{\mathrm{e}}\right)\left(e^{x}-1\right)(xe^{x})^{-1}
\cdot T_\textrm{CMB}$ 
\citep[e.g.,][]{Bonamente2007}. 
Thus for a cluster observed at a given frequency, 
the $\Delta I_\textrm{SZE}$ (Eq.~(\ref{eq:sz-intensity})) and $\Delta T_\textrm{SZE}$ are
equivalent measures of the Compton-$y$ parameter.
In Table~\ref{tab:SZ-obs},
we include the values of the central thermodynamic temperature decrement $\Delta T_0$
that correspond to the $\Delta I_0$ based on \textit{Chandra}.

To obtain an overall measure of the thermal energy content in a cluster,
we computed the integrated Compton-$y$ parameter $Y_{2500}$,
which is the Compton-$y$ integrated from its center out to the projected radius $r_{2500}$,
\begin{equation}
\label{eq:inensity2Y}
\begin{split}
	Y_{2500} & \equiv\int_{\Omega_{2500}}y\: d\Omega\\
 & =\frac{2\pi\Delta I_{0}}{I_\textrm{CMB}\, g(x,T_{\mathrm{e}})}\int_{0}^{r_{2500}/D_{\mathrm{A}}}\left(1+\frac{\theta^{2}}{\theta_{\mathrm{c}}^{2}}\right)^{(1-3\beta)/2}\theta d\theta,
\end{split}
\end{equation}
where $\Omega$ is the solid angle of the integrated patch
and $\Omega_{2500}$ is the total value covered within radius $r_{2500}$.
The integrated Compton-$y$ parameter has been shown to be a more
robust quantity than the central value of Compton-$y$ for observational tests,
because it is less dependent on the model of gas distribution used for the analysis \citep{Benson2004}.
In addition, integrating the Compton-$y$ out to a large projected radius diminishes (though does not completely remove)
effects resulting from the presence of strong entropy features in the central regions of clusters \citep{McCarthy2003a}.
Table \ref{tab:SZ-obs} summarizes our derived values of $Y_{2500}$, adopting the parameters based on \textit{Chandra}. 
In Section~\ref{sec:scalings},
the $Y_{2500}$ derived from both X-ray parameter sets will be considered for its scaling relationship
with $T_{\mathrm{e}}$, $M_{2500}$ and $L_{\mathrm{X}}$.

Although using X-ray data to determine the shapes of cluster SZE
profiles is a common strategy in SZE analysis, it has been shown that
this will bias the results of fitted parameters due to the assumption
of isothermality of a $\beta$-model
\citep[e.g.~][]{Komatsu2001,Hallman2007}. In Section~\ref{sec:dandc}
we will further discuss this issue, and apply a simple correction to
our results based on the work of \citet{Hallman2007}.

AMiBA is one of the first instruments to provide 3-mm SZ detections of the
cluster targets, expanding our knowledge of the SZE spectra for clusters.
Table~\ref{tab:SZ-obs} compares our results for $Y_{2500}$ at 94~GHz 
with results at other frequencies: the BIMA/OVRO results at 30~GHz \citep{McCarthy2003b,Morandi2007}, 
and the SuZIE~II results at 145~GHz \citep{Benson2004}. 
We have converted the BIMA/OVRO values of integrated Compton-$y$ parameter, 
$y_{2500}$ \citep{Morandi2007}, to our $Y_{2500}$ using
$Y_{2500}=y_{2500}\, (F-4)\, xe^{x}/\left[(e^{x}-1)\, I_0\,
g(x,T_{\mathrm{e}})\right]$ where $I_0\equiv2\left(k_{\mathrm{B}}T_{\mathrm{CMB}}\right)^3 (h\mathrm{c})^{-2}$, and $x$, $F$, and $g(x,T_{\mathrm{e}})$ are defined in Eq.~(\ref{eq:freqfactor}) with frequency $\nu=30$~GHz.
For SuZIE~II, $Y_{2500}$ is
obtained from $Y_{2500}=S\left(r_{2500}\right)/\left[I_{\mathrm{CMB}}\, g(x,T_{\mathrm{e}})\right]$, where
$S\left(r_{2500}\right)$ is the integrated SZ flux defined in
\citet{Benson2004}, and $I_{\mathrm{CMB}}$ and $g(x,T_{\mathrm{e}})$ are calculated at $\nu=30$~GHz. 
All three sets of results are based on
reconstruction of the gas profile of the clusters using an isothermal
$\beta$-model, and all include the relativistic correction in their
estimates of $Y_{2500}$. Our results are consistent with those from
BIMA/OVRO except for A1995, and can be seen to be generally lower than
those from SuZIE~II. 

Since $Y_{2500}$ is relatively insensitive to cluster morphology and
the thermal structure of the intracluster medium, the difference
between the results for A1995 could mostly come from the different SZ
techniques. The AMiBA data led to a signal-to-noise ratio of about 6,
based on an integration time of about 5.5 hours. A1995 is a
relatively low-mass, cool, cluster and hence produces a smaller SZE
than the other clusters in our sample. This causes the statistical
significance of the different values for $Y_{2500}$ to be low, and that
differing residual contamination by point sources and primary CMB
fluctuations \citep{Liu2008} could be an important factor. Further
multi-frequency studies of A1995 are needed to improve the 
result for the SZE of this cluster.

\begin{deluxetable*}{l|cc|ccc}
\tablewidth{0pt}
\tablecaption{Parameters of AMiBA clusters derived from SZE observations
\label{tab:SZ-obs}} 
\tablehead{
        & $\Delta I_{0}$ & $\Delta T_{0}$  & \multicolumn{3}{c}{$Y_{2500}$ ($10^{-10}$ sr)} \\
Cluster & ($10^{5}$Jy/sr)\tablenotemark{a} & (mK) & BIMA/OVRO & AMiBA & SuZIE II
}
\startdata 
A1689  & $-2.36\pm0.71$ & $-0.40\pm0.12$ & $2.17\pm0.14$ & $2.82\pm0.86$ & $4.65_{-0.51}^{+0.61}$\\
A1995  & $-3.19\pm1.23$ & $-0.54\pm0.21$ & $0.71\pm0.06$ & $1.49\pm0.58$ & --\\
A2142  & $-2.09\pm0.36$ & $-0.35\pm0.06$ & -- & $13.44\pm2.40$ & --\\
A2163  & $-3.64\pm0.61$ & $-0.62\pm0.10$ & $5.53\pm0.41$ & $6.61\pm1.38$ & $5.50_{-0.70}^{+0.76}$\\
A2261  & $-2.59\pm0.90$ & $-0.44\pm0.15$ & $1.51\pm0.18$ & $1.72\pm0.64$ & $4.46_{-0.94}^{+1.70}$\\
A2390  & $-2.85\pm0.77$ & $-0.48\pm0.13$ & -- & $3.12\pm0.98$ & $3.69_{-0.57}^{+0.56}$
\enddata

\tablecomments{The integrated Compton parameters $Y_{2500}$ measured
by AMiBA (94~GHz) are compared with results from BIMA/OVRO
\citep[30~GHz;][]{McCarthy2003b,Morandi2007} 
and SuZIE~II \citep[145~GHz; deduced from][]{Benson2004}. 
The central SZE intensity $\Delta I_{0}$, its corresponding
thermodynamic temperature decrement $\Delta T_{0}$, and the AMiBA
$Y_{2500}$ were derived using the \textit{Chandra}-based parameters. 
Isothermal $\beta$-models are used in all three sets of observation
to reconstruct the gas profile of clusters and derive $Y_{2500}$. The
relativistic correction $\delta_\textrm{rel}\left(x,T_{\mathrm{e}}\right)$
in Eq.~(\ref{eq:relativistic}) is also taken into account in all
three cases.
Errors are given at the $68.3 \%$ confidence level.
}
\tablenotetext{a}{Central SZ intensities are given by \citet{Liu2008}.}
\end{deluxetable*}

\section{SCALING RELATIONS}
\label{sec:scalings}

\subsection{Theoretical Expectations}
\label{sec:sub:scaletheory}

In the context of the self-similar model, if assuming hydrostatic
equilibrium and an isothermal distribution of baryons in the spherically-collapsed
DM halo, it can be shown that there are simple power-law scaling relations
between the SZE and X-ray quantities.
Specifically there are simple relations between the integrated Comptonization and the gas temperature
$T_{\mathrm{e}}$, the cluster total mass $M_{tot}$ and the bolometric
X-ray luminosity $L_{\mathrm{X}}$,
\begin{eqnarray}
YD_{\mathrm{A}}^{2}& \propto  T_{\mathrm{e}}^{5/2}E(z)^{-1} ,\\
YD_{\mathrm{A}}^{2}& \propto  M_{tot}^{5/3}E(z)^{2/3}, \\
YD_{\mathrm{A}}^{2}& \propto  L_{\mathrm{X}}^{5/4}E(z)^{-9/4},
\end{eqnarray}
where $E^{2}(z)=\Omega_{\mathrm{M}}(1+z)^{3}+\Omega_{\Lambda}+\Omega_{\mathrm{k}}(1+z)^{2}$ \citep{Morandi2007}.
We note that these scaling relations assume that the fraction of the cluster mass
present as gas, $f_{{\rm gas}}$, is a constant. 
\citet{Bonamente2007} found no significant scatter of $f_{gas}$ in their results.
Nevertheless, recent X-ray work in observations (e.g.~\citealp{Vikhlinin2006a}) and simulations
\citep{Kravtsov2005} suggest that some variation may be expected.

Following standard method (e.g.~\citealp{Press2002}), we perform a linear
least-squares fitting in $\log_{10}$ space, $\log_{10}(y)=A+B\log_{10}(x)$,
taking account of errors in both $x$ and $y$, to estimate the normalization $A$
and power law index $B$ of each scaling relation. 
The $\chi^{2}$ statistic is defined as 
\begin{equation}
\chi^{2}=\sum\frac{\left(\log_{10}(y_{i})-A-B\log_{10}(x_{i})\right)^{2}}{\left(\sigma_{y_{i}}\log_{10}(e)/y_{i}\right)^{2}+\left(B\sigma_{x_{i}}\log_{10}(e)/x_{i}\right)^{2}}\quad,\label{eq:chi-square}
\end{equation}
and is minimized as in \citet[Eq.~(13)]{Benson2004}. $\sigma_{y_{i}}$
and $\sigma_{x_{i}}$ for the sample points are obtained from the upper
and lower uncertainties around the best-fit values as $\sigma=\left(\sigma^{+}+\sigma^{-}\right)/2$.
The number of degrees of freedom (d.o.f.) is $N-2$ with $N$ equal to the total number of clusters in the sample.
$1\sigma$ errors in $A$ and $B$ are determined by projecting the
$\Delta\chi^{2}=1$ contour on each coordinate axis.

\subsection{Derived Observational Results}
\label{sec:sub:results}
The results of fitting $\log_{10}(y)=A+B\log_{10}(x)$ for each scaling
relation are summarized in Table~\ref{tab:srline}. Figures~\ref{fig:YvsTe},
\ref{fig:YvsM2500}, and~\ref{fig:YvsLx} show our sample of six
clusters and the best-fitting scaling relations. 
Each figure shows the
scaling results based on the \textit{Chandra} and the \textit{ASCA/ROSAT}
parameters, for comparison.
Five of the six clusters in our sample are cooling-core (CC) clusters;
the exception is cluster A2163, which is of non-cooling core (NCC) type 
\citep{Myers1997,Allen2000,McCarthy2003b,Morandi2007}.

\begin{deluxetable*}{cccccccccc}
\tabletypesize{\scriptsize}
\tablewidth{0pt}
\tablecaption{Scaling relations from X-ray and AMiBA SZE data
\label{tab:srline}} 
\tablehead{ 
 \multicolumn{2}{c}{$\log_{10}(y)=A+B\log_{10}(x)$} & & \multicolumn{3}{c}{\textit{Chandra}} & & \multicolumn{3}{c}{\textit{ASCA/ROSAT}}\\
\cline{1-2} \cline{4-6}  \cline{8-10}\\
 $x$ & $y$ & & $A$ & $B$ & $\chi_{\mathrm{min}}^{2}$(d.o.f.) & & $A$ & $B$ & $\chi_{\mathrm{min}}^{2}$(d.o.f.)
}
\startdata
$T_{\mathrm{e}}$/keV & $Y_{2500}D_{\mathrm{A}}^{2}E(z)/{\rm Mpc^{2}}$ &
& $-5.94_{-0.72}^{+0.67}$ & $2.28_{-0.68}^{+0.73}$ & $1.43(4)$ &
& $-5.97_{-0.78}^{+0.67}$ & $2.21_{-0.64}^{+0.74}$ & $3.36(4)$\\
 $M_{2500}/10^{14}M_{\sun}$ & $Y_{2500}D_{\mathrm{A}}^{2}E(z)^{-2/3}/{\rm Mpc^{2}}$ &
& $-4.82_{-0.59}^{+0.39}$ & $1.71_{-0.64}^{+1.01}$ & $1.82(4)$  &
& $-5.03_{-0.60}^{+0.43}$ & $1.90_{-0.61}^{+0.83}$ & $1.95(4)$\\
 $L_{\mathrm{X}}/10^{45}{\rm erg\, s^{-1}}$ & $Y_{2500}D_{\mathrm{A}}^{2}E(z)^{9/4}/{\rm Mpc^{2}}$ &
& $-4.05_{-0.18}^{+0.18}$ & $0.77_{-0.32}^{+0.32}$ & $6.15(3)$  &
& $-4.69_{-0.28}^{+0.28}$ & $1.11_{-0.29}^{+0.28}$ & $1.69(3)$
\enddata

\tablecomments{The $Y_{2500}-L_{\mathrm{X}}$ fit uses only five
clusters, omitting A2142 in the \textit{Chandra} set and omitting
A1995 in the \textit{ASCA/ROSAT} set. $\chi_{\mathrm{min}}^{2}$
gives the minimum value of $\chi^{2}$, as defined in
Eq.~(\ref{eq:chi-square}), with the corresponding number of degrees of
freedom (d.o.f.).
Errors are given at the $68.3 \%$ confidence level.
}
\end{deluxetable*}

\subsubsection{The $Y_{2500}$ --$T_{\mathrm{e}}$ relation}
\label{sec:sub:Y2500T}

Our results for the power law index $B$ from both sets of X-ray
parameters agree with the self-similar model $B=2.5$ at the $1\sigma$ level.
They are also consistent with the values of 
$B=2.37\pm0.23$ from BIMA/OVRO \citep{Bonamente2007}, 
$B=2.21\pm0.41$ from SuZIE~II \citep{Benson2004}, 
and $B=2.64\pm0.28$ (CC+NCC sample) and $B=2.74\pm0.23$ (CC sample only) from \citet{Morandi2007}. 

To compare the normalization in scaling relations in the same analytic form and units, 
we convert the SuZIE~II normalizations to
$A=A^{\prime} -\log_{10}\left[I_0\, g(x,T_{\mathrm{e}}) x^{3}(e^{x}-1)^{-1}\right]$, 
where $I_{0}\equiv2\left(k_\textrm{B}T_\textrm{CMB}\right)^{3}\left(h\mathrm{c}\right)^{-2}$, $x$ is calculated at the SuZIE~II observing frequency of 145~GHz,
and the primes stand for the power indices or normalizations from the
references that we compare. For normalizations of \citet{Morandi2007},
$A=A^{\prime}-\log_{10}\left(10^{-8}\,I_{0}\right)
-B^{\prime}\log_{10}\left(7\right)$, following the same convention as in SuZIE~II case.
Our values for the normalization, $A$, in both sets are consistent within $1\sigma$ with the values 
$A=-6.24\pm0.22$ from BIMA/OVRO \citep{Bonamente2007}, 
$-6.64\lesssim A\lesssim-5.82$ from SuZIE~II \citep{Benson2004}, 
and $-6.67\lesssim A\lesssim-6.14$ for a combined CC+NCC sample and $-6.80\lesssim A\lesssim-6.35$ for a CC-only
sample from \citet{Morandi2007}.

\begin{figure}
	\plotone{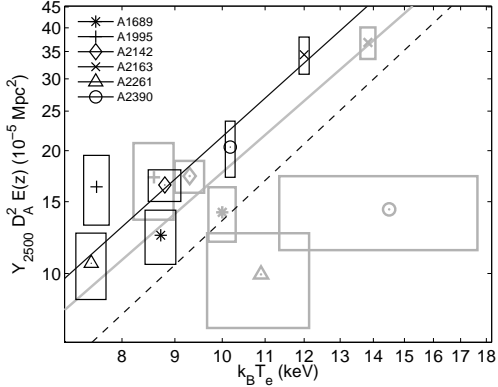}
  \caption{\small 
	The scaling relation between $Y_{2500}$ and
	$T_{\mathrm{e}}$. Those in solid black show the
	\textit{Chandra}-based results. Those in gray show the
	\textit{ASCA/ROSAT}-based results. 
	Six clusters are labeled as indicated by the legend,
	with errors represented by the boxes.
	The lines are the best-fit power-law relations. 
	The black dashed line is the best fit from \citet{Bonamente2007}
	for comparison.
	\label{fig:YvsTe}}
\end{figure}

\begin{figure}
	\plotone{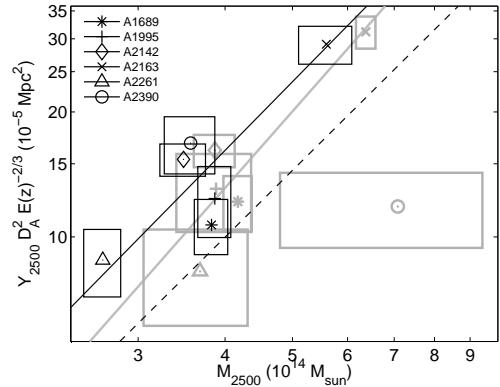}
  \caption{\small 
	The scaling relation between $Y_{2500}$ and $M_{2500}$.
	Symbols and colors are as defined in Fig.~\ref{fig:YvsTe}.
	The black dashed line refers 		to the best-fit from \citet{Bonamente2007} for comparison.
	\label{fig:YvsM2500}}
\end{figure}

Figure~\ref{fig:YvsTe} shows that the scaling relation based on
the \textit{ASCA/ROSAT} parameters has a lower normalization than the \textit{Chandra}-based relation
due to the systematically higher temperatures.
The scaling relation is not well confined by the \textit{ASCA/ROSAT},
partly due to larger errors and partly due to the fact
the scaling is defined by only a scatter of the five CC clusters and the single NCC cluster A2163.
As briefly mentioned in Section~\ref{sec:xrpars}, if A2142
is removed from the \textit{Chandra} set, since its model is somewhat
inconsistent with the others, the change on the scaling relation is
less than $5\%$ because A2142 lies close to the best-fit line.

\subsubsection{The $Y_{2500}$ -- $M_{2500}$ relation}
\label{sec:sub:YMtot}
The power-law index $B$ based on both sets of X-ray parameters
are consistent with the self-similar model prediction of $B=1.67$.
Our results also agree with the values of
$B=1.66\pm0.20$ from BIMA/OVRO \citep{Bonamente2007}, 
and $B=1.48\pm0.39$ (CC+NCC sample) and $B=1.56\pm0.29$ (CC only) from \citet{Morandi2007}.
Our normalization agrees with the value 
$A = -5.0 \pm 3.0$ from BIMA/OVRO \citep{Bonamente2007} 
and is consistent with the ranges 
$-5.09\lesssim A\lesssim-4.63$ (CC+NCC samples) and 
$-5.36\lesssim A\lesssim-4.91$ (CC only) given by \citet{Morandi2007}.
Here we convert the normalizations from other studies for comparison, such as
$A=A^{\prime}+14B^{\prime}$ for BIMA/OVRO, and
$A=A^{\prime}+\left(B^{\prime}-5/3\right)\log_{10}\overline{E(z)}-\log_{10}(10^{-8} I_{0})$
for \citet{Morandi2007}, 
where $\overline{E(z)}$ is the mean $E(z)$ averaged over all AMiBA clusters.

Several analytical and numerical studies demonstrate that the
integrated SZE signal, $Y_{2500}$ in our case, as a measure of the total pressure
of inter-cluster medium is an excellent proxy for cluster total mass
\citep{Silva2004,Motl2005,Nagai2006,Hallman2007}. 
If this relationship could be measured to high precision at low redshifts, 
it could then be used to determine the masses of
high-redshift SZE clusters and test cosmological models.

\subsubsection{The $Y_{2500}$ -- $L_{\mathrm{X}}$ relation}
\label{sec:sub:yl}
\begin{figure}
	\plotone{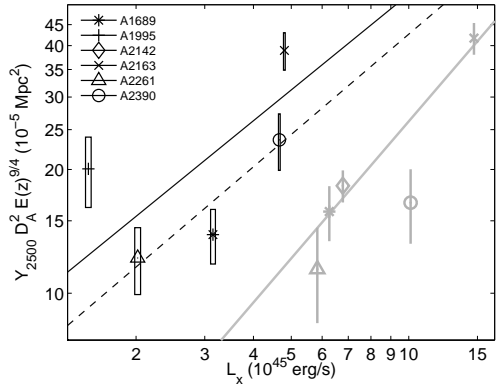}
  \caption{\small 
	The scaling relations between $Y_{2500}$ and
	$L_{\mathrm{X}}$. 
	Symbols and colors are as defined in Fig.~\ref{fig:YvsTe}.
	The black dashed line shows the best-fit relation from
	\citet{Morandi2007} for comparison.
	The luminosities from the \textit{ASCA/ROSAT} set
	 are given without errors.
	\label{fig:YvsLx}}
\end{figure}

The power-law index for the $Y_{2500}-L_{\mathrm{X}}$ relation
based on the \textit{Chandra} set (after omitting A2142) is about $1.5\sigma$
lower than the theoretical value $B=1.25$, but is consistent with
the results $B=0.81\pm0.07$ (for CC+NCC sample) and $B=0.91\pm0.11$ (for CC sample) given by \citet{Morandi2007}.
A low power-law index has also been observed in numerical simulations that include
cooling or preheating processes \citep[$Y\propto L_{\mathrm{X}}$]{Silva2004}.
The systematically lower power-law index relative to the self-similar model predication
seems to imply that the relation between the SZE signals and X-ray luminosities is
more sensitive to the radiative content outside 100~kpc than other scaling relations.
However, none of the data points in Fig.~\ref{fig:YvsLx} lies close to the best-fit line
and the measure of goodness-of-fit, $\chi_{\mathrm{min}}^{2}/\mathrm{d.o.f.}$, is large (see Tab.~\ref{tab:srline}). 
On the other hand, the normalization agrees with the result
$-4.35\lesssim A\lesssim-4.01$ for CC+NCC sample within $1\sigma$ 
and is broadly consistent with $-4.47\lesssim A\lesssim-4.36$ for CC sample \citep{Morandi2007}, through the conversion of
$A=A^{\prime}-(5/4-B^{\prime})\log_{10}\overline{E(z)}-\log_{10}\left(10^{-8}I_{0}\right)+B^{\prime}$.

The power-law index and normalization based on the \textit{ASCA/ROSAT} set are consistent with the self-similar model.
They also agree with the result based on the CC sample given by \citet{Morandi2007} within $1\sigma$, 
but only marginally consistent with those based on the CC+NCC sample. 
We observed, by comparing the values of different models in \citet{Allen2000}, that the additional component compensating the cooling flow emission in Model~C will generally reduce the bolometric luminosities. If there is residual luminous emission, as \citet{Morandi2007} remarked that CC clusters systematically have larger luminosities than NCC ones 
even if the cooling cores have been handled, it would bias high the power (slope) and bias low the normalization (interception) shown in Fig.~\ref{fig:YvsLx} in the sense of shifting the CC clusters to higher  $L_{\mathrm{X}}$. This is a possible interpretation of the discrepancy for the CC+NCC sample, but again the scaling relation is not well defined, essentially by a scatter of CC cluster and a NCC cluster outside the scatter.

\section{DISCUSSIONS}
\label{sec:dandc}

In the three figures of scaling relations we see that the
\textit{Chandra}-based relations are generally better fits than the 
\textit{ASCA/ROSAT}-based relations, with smaller
$\chi_{\mathrm{min}}^{2}$ and smaller errors on each data point. Although the
$Y-L_{\mathrm{X}}$ relation based on the \textit{Chandra} set has a larger
scatter, there are no errors available for the luminosities of the
\textit{ASCA/ROSAT} set. Among clusters in the \textit{ASCA/ROSAT}
set, A2261 and A2390 have X-ray parameters of poor quality. A2390
seems to have a biased-high gas temperature or a systematically low $Y_{2500}$. 
A high temperature would lead to a high total mass
based on the hydrostatic equilibrium equation, and would
similarly increase the luminosity, which is related to the gas temperature 
as $L_{\mathrm{X}}\propto T_{\mathrm{e}}^{2}$ \citep{Morandi2007}.

\begin{deluxetable*}{ccccc}
\tablewidth{0pt}
\tablecaption{Scaling relations $\log_{10}(y)=A+B\log_{10}(x)$ corrected for isothermal $\beta$-model
\label{tab:beta-correction}} 
\tablehead{ 
 $x$ & $y$ & $A$ & $B$ & $\chi_{\mathrm{min}}^{2}$(d.o.f.)}
\startdata
 $T_{\mathrm{e}}/keV$ & $Y_{2500}D_{\mathrm{A}}^{2}E(z)/{\rm Mpc^{2}}$ & 
 $-6.16_{-0.74}^{+0.68}$ & $2.31_{-0.70}^{+0.75}$ & $1.74(4)$\\
 $M_{2500}/10^{14}M_{\sun}$ & $Y_{2500}D_{\mathrm{A}}^{2}E(z)^{-2/3}/{\rm Mpc^{2}}$ & 
 $-4.82_{-0.43}^{+0.30}$ & $1.68_{-0.59}^{+0.87}$ & $1.73(4)$\\
 $L_{\mathrm{X}}/10^{45}{\rm erg\, s^{-1}}$ & $Y_{2500}D_{\mathrm{A}}^{2}E(z)^{9/4}/{\rm Mpc^{2}}$ & 
 $-4.24_{-0.18}^{+0.18}$ & $0.79_{-0.32}^{+0.32}$ & $6.72(3)$
\enddata
\tablecomments{Errors are given at the $68.3 \%$ confidence level.}
\end{deluxetable*}

Analytical and numerical studies reveal the fundamental incompatibility
between $\beta$-model fits to X-ray surface brightness profiles and
those done with SZE profiles
\citep[e.g.~][]{Komatsu2001,Hallman2007}. Both X-ray and SZE fitted 
model parameters are biased due to the isothermal assumption, since
the X-ray surface brightness and SZE Compton-$y$ parameter have 
different dependences on the cluster temperature profiles. This
will generate an inconsistency in the model parameters based on
isothermal $\beta$-model fits. Since observational SZE radial profiles
are in short supply, X-ray driven parameters are often used to
constrain the profile shape in SZE analysis, 
consequently leading to a bias in the derived values of cluster mass or Comptonization parameter.  

To remedy this problem, we followed \citet{Hallman2007}.
Instead of re-fitting by the universal temperature profile
proposed by \citet{Hallman2007}, we simply modify our values of
$\beta$, $r_{\mathrm{c}}$ and $Y_{2500}$ by the ratios 
between the values fitted from X-ray data on an isothermal
$\beta$-model, and the `true' values obtained from the
simulation. 
We then re-calculate the scaling relations and these corrected results
are summarized in Table~\ref{tab:beta-correction}. 
It is clear that the introduction of
correction still keeps the scaling relations consistent with the uncorrected results, 
and the previous arguments and discussions are still valid. 
The scaling relations seem to be insensitive to this correction. As
\citet{Hallman2007} discovered in their study of the $Y-M_{gas}$ relation, 
for example, the correlated changes in a $\beta$-model
due to the definition of projected radius ($r_{2500}$ in our case) 
tend to cause compensating changes in the scaling relation.

We are aware that the entropy floor present in the cores of clusters
could give rise to deviations from self-similar
scalings (see e.g., \citet{McCarthy2003a,McCarthy2003b}). X-ray
observations have shown that scaling relations between several cluster
observables deviate from the self-similar prediction, and it has
been found that heating and cooling act in a similar manner by raising
the mean entropy of the intracluster gas and, in some cases,
establishing a core in the entropy profile. In \citet{McCarthy2003a}
it was observed that the injection of excess entropy (preheating) will
increase the temperature and reduce the gas pressure in the central
regions of clusters, especially for the low-mass clusters. The scaling
relations between the central value of the Compton-$y$ parameter,
$y_{0}$, and the gas temperature or the total mass are most sensitive
to the presence of excess entropy, and tend to develop larger power
indices. Scaling relations involving the integrated Compton-$y$
parameter $Y$ show similar behaviors but are less sensitive, since the
integration tends to reduce the effect of the entropy contribution from
the cores.

\section{CONCLUSIONS}
\label{sec:conclusions}

In understanding cluster physics and cosmic evolution,
the study of scaling relations for galaxy clusters is becoming more important 
since SZE observations such as those from the Sunyaev--Zel'dovich Array (SZA) and the South Pole
Telescope (SPT) will detect clusters which are beyond X-ray
detection limits, for which SZE scaling relations provide the
first indications of cluster properties.
In addition, deviations between the theoretical and observational results of the scaling
relations can serve to examine the non-gravitational processes
in the formation of clusters, which are not well understood at present.

As one of the few SZE instruments working at 3-mm, the
AMiBA experiment observed six Abell clusters during 2007. The derived
integrated Compton-$y$ parameters, $Y_{2500}$, are compared to other
observations at different frequencies, as summarized in Table~\ref{tab:SZ-obs}.
Our results are consistent with those from BIMA/OVRO, but appear to
show lower Comptonizations than those from SuZIE~II. We have also
investigated the three scaling relations between $Y_{2500}$ and the X-ray
spectroscopic temperatures, total masses within $r_{2500}$, and
bolometric X-ray luminosities. 
Our results for the scaling relations are summarized in Table~\ref{tab:srline}.

Our power-law indices for the three scaling relations are broadly consistent
with the self-similar model and observational results in the literature,
except for that the $Y_{2500}-L_{\mathrm{X}}$ relation based on
\textit{Chandra}-derived parameters has a slope lower than the expectation of the self-similar model,
and is sensitive to the chosen set of X-ray parameters and the treatment of cooling cores.
These discrepancies might indicate either exotic properties for clusters or hidden flaws in our SZE data,
although the scatter is still large, about a factor of two in the integrated
Compton-$y$ relative to the fit line. 
The agreement between the normalizations found by different workers for our three scaling relations seems
to support the idea that there is no strong scatter in the gas fraction
(see Sec.~\ref{sec:sub:scaletheory}).

In conclusion, the agreement between our results and those from the literature provides
not only confidence for this project but also supports to our understanding
of galaxy clusters. For AMiBA, significant improvements are expected
following the expansion to a 13-element configuration
with 1.2-m antennas \citep[AMiBA13]{Ho2008}, 
which will provide better resolution and higher sensitivity.
The capability of resolving SZE clusters
will make it possible to measure the cluster profiles independent of the X-ray data \citep{Molnar2008}
and to estimate the properties of the clusters which currently do not have good X-ray data.

\acknowledgements
We thank the Ministry of Education, the National Science Council,
the Academia Sinica, and National Taiwan University for their support of this project. 
We thank the Smithsonian Astrophysical Observatory for hosting the AMiBA project staff at the SMA Hilo Base Facility. 
We thank the NOAA for locating the AMiBA project on their site on Mauna Loa. 
We thank the Hawaiian people for allowing astronomers to work on their mountains in order to study the Universe.
We are grateful for computing support from the National Center for High-Performance Computing, Taiwan.
This work is also supported by National Center for Theoretical Science, and
Center for Theoretical Sciences, National Taiwan University for J.H.P.~Wu.
We appreciate the extensive comments on this article from Katy Lancaster. 
Support from the STFC for MB is also acknowledged.

\bibliographystyle{hapj}
\bibliography{/home/locutus/AMiBA/bibs/sz,/home/locutus/AMiBA/bibs/x-ray,/home/locutus/AMiBA/bibs/others,/home/locutus/AMiBA/bibs/cmb,/home/locutus/AMiBA/bibs/lss}

\begin{thebibliography}{47}
\expandafter\ifx\csname natexlab\endcsname\relax\def\natexlab#1{#1}\fi

\bibitem[{Allen(2000)}]{Allen2000}
Allen, S.~W. 2000, \mnras, 315, 269

\bibitem[{Allen {et~al.}(2001)Allen, Ettori, \& Fabian}]{Allen2001}
Allen, S.~W., Ettori, S., \& Fabian, A.~C. 2001, \mnras, 324, 877

\bibitem[{Allen \& Fabian(1998)}]{Allen1998a}
Allen, S.~W., \& Fabian, A.~C. 1998, \mnras, 297, L57

\bibitem[{Benson {et~al.}(2004)Benson, Ade, Bock, Ganga, Henson, Thompson, \&
  Church}]{Benson2004}
Benson, B., Ade, P., Bock, J., Ganga, K., Henson, C., Thompson, K., \& Church,
  S. 2004, \apj, 617, 829

\bibitem[{B\"{o}hringer {et~al.}(1998)B\"{o}hringer, Tanaka, Mushotzky, Ikebe,
  \& Hattori}]{Bohringer1998}
B\"{o}hringer, H., Tanaka, Y., Mushotzky, R.~F., Ikebe, Y., \& Hattori, M.
  1998, \aap, 334, 789

\bibitem[{Bonamente {et~al.}(2008)Bonamente, Joy, LaRoque, Carlstrom, Nagai, \&
  Marrone}]{Bonamente2007}
Bonamente, M., Joy, M., LaRoque, S.~J., Carlstrom, J.~E., Nagai, D., \&
  Marrone, D.~P. 2008, \apj, 675, 106, 0708.0815

\bibitem[{Bonamente {et~al.}(2006)Bonamente, Joy, Roque, Carlstrom, Reese, \&
  Dawson}]{Bonamente2006}
Bonamente, M., Joy, M., Roque, S.~L., Carlstrom, J., Reese, E., \& Dawson, K.
  2006, \apj, 647, 25

\bibitem[{Cavaliere \& Fusco-Femiano(1976)}]{CF1976}
Cavaliere, A., \& Fusco-Femiano, R. 1976, \aap, 49, 137

\bibitem[{Challinor \& Lasenby(1998)}]{Challinor1998}
Challinor, A.~D., \& Lasenby, A.~N. 1998, \apj, 499, 1

\bibitem[{Chen {et~al.}(2009)Chen, Li, Hwang, Jiang, Altamirano, Chang, Chang,
  Chang, Chiueh, Chu, Han, Huang, Kesteven, Kubo, Martin-Cocher, Oshiro,
  Raffin, Wei, Wang, Wilson, Ho, Huang, Koch, Liao, Lin, Liu, Molnar, Nishioka,
  Umetsu, Wang, \& Wu}]{Chen2008}
Chen, M.-T. {et~al.} 2009, \apj, 694, 1664, 0902.3636

\bibitem[{da~Silva {et~al.}(2004)da~Silva, Kay, Liddle, \& Thomas}]{Silva2004}
da~Silva, A.~C., Kay, S.~T., Liddle, A.~R., \& Thomas, P.~A. 2004, \mnras, 348,
  1401

\bibitem[{Grego {et~al.}(2001)Grego, Carlstrom, Reese, Holder, Holzapfel, Joy,
  Mohr, \& Patel}]{Grego2001}
Grego, L., Carlstrom, J.~E., Reese, E.~D., Holder, G.~P., Holzapfel, W.~L.,
  Joy, M.~K., Mohr, J.~J., \& Patel, S. 2001, \apj, 552, 2

\bibitem[{Hallman {et~al.}(2007)Hallman, Burns, Motl, \& Norman}]{Hallman2007}
Hallman, E.~J., Burns, J.~O., Motl, P.~M., \& Norman, M.~L. 2007, \apj, 665,
  911

\bibitem[{Ho {et~al.}(2009)Ho, Wu, Huang, Koch, Liao, Lin, Liu, Molnar,
  Nishioka, Umetsu, Wang, Altamirano, Chang, Chang, Chang, Chen, Han, Huang,
  Hwang, Jiang, Kesteven, Kubo, Li, Martin-Cocher, Oshiro, Raffin, Wei, \&
  Wilson}]{Ho2008}
Ho, P.~T.~P. {et~al.} 2009, \apj, 694, 1610, 0810.1871

\bibitem[{Kaiser(1986)}]{Kaiser1986}
Kaiser, N. 1986, \mnras, 222, 323

\bibitem[{Koch {et~al.}(2009{\natexlab{a}})Koch, Wu, Ho, Huang, Liao, Lin, Liu,
  Molnar, Nishioka, Umetsu, Wang, Altamirano, Chang, Chang, Chang, Chen, Han,
  Huang, Hwang, Jiang, Kesteven, Kubo, Li, Martin-Cocher, Oshiro, Raffin, Wei,
  \& Wilson}]{Koch2008a}
Koch, P.~M. {et~al.} 2009{\natexlab{a}}, \apj, 694, 1670

\bibitem[{Koch {et~al.}(2009{\natexlab{b}})Koch, Wu, P.Ho, Huang, Liao, Lin,
  Liu, Molnar, Nishioka, Umetsu, Wang, Altamirano, Chang, Chang, Chang, Chen,
  Han, Huang, Hwang, Jiang, Kesteven, Kubo, Li, Martin-Cocher, Oshiro, Raffin,
  Wei, \& Wilson}]{Koch2008}
------. 2009{\natexlab{b}}, \apj, submitted

\bibitem[{Komatsu \& Seljak(2001)}]{Komatsu2001}
Komatsu, E., \& Seljak, U. 2001, \mnras, 327, 1353

\bibitem[{Kravtsov {et~al.}(2005)Kravtsov, Nagai, \& Vikhlinin}]{Kravtsov2005}
Kravtsov, A.~V., Nagai, D., \& Vikhlinin, A.~A. 2005, \apj, 625, 588

\bibitem[{Lancaster {et~al.}(2005)Lancaster, Genova-Santos, Falc\`{o}n,
  Grainge, Guti\`{e}rrez, Kneissl, Marshall, Pooley, Rebolo, Rubi\~{n}o-Martin,
  Saunders, Waldram, \& Watson}]{Lancaster2005}
Lancaster, K. {et~al.} 2005, \mnras, 359, 16, astro-ph/0405582

\bibitem[{LaRoque {et~al.}(2006)LaRoque, Bonamente, Carlstrom, Joy, Nagai,
  Reese, \& Dawson}]{LaRoque2006}
LaRoque, S.~J., Bonamente, M., Carlstrom, J.~E., Joy, M.~K., Nagai, D., Reese,
  E.~D., \& Dawson, K.~S. 2006, \apj, 652, 917

\bibitem[{Lin {et~al.}(2009)Lin, Wu, P.Ho, Huang, Koch, Liao, Liu, Molnar,
  Nishioka, Umetsu, Wang, Altamirano, Chang, Chang, Chang, Chen, Han, Huang,
  Hwang, Jiang, Kesteven, Kubo, Li, Martin-Cocher, Oshiro, Raffin, Wei, \&
  Wilson}]{Lin2008}
Lin, K.-Y. {et~al.} 2009, \apj, 694, 1629, 0902.2437

\bibitem[{Liu {et~al.}(2009)Liu, Wu, P.Ho, Huang, Koch, Liao, Lin, Molnar,
  Nishioka, Umetsu, Wang, Altamirano, Chang, Chang, Chang, Chen, Han, Huang,
  Hwang, Jiang, Kesteven, Kubo, Li, Martin-Cocher, Oshiro, Raffin, Wei, \&
  Wilson}]{Liu2008}
Liu, G.-C. {et~al.} 2009, \apj, submitted

\bibitem[{Markevitch {et~al.}(2000)Markevitch, Ponman, Nulsen, Bautz, Burke,
  David, Davis, Donnelly, Forman, Jones, Kaastra, Kellogg, Kim, Kolodziejczak,
  Mazzotta, Pagliaro, Patel, Speybroeck, Vikhlinin, Vrtilek, Wise, \&
  Zhao}]{Markevitch2000}
Markevitch, M. {et~al.} 2000, \apj, 541, 542

\bibitem[{Mather {et~al.}(1999)Mather, Fixsen, Shafer, Mosier, \&
  Wilkinson}]{Mather1999}
Mather, J.~C., Fixsen, D.~J., Shafer, R.~A., Mosier, C., \& Wilkinson, D.~T.
  1999, \apj, 512, 511

\bibitem[{McCarthy {et~al.}(2002)McCarthy, Babul, \& Balogh}]{McCarthy2002}
McCarthy, I.~G., Babul, A., \& Balogh, M.~L. 2002, \apj, 573, 515

\bibitem[{McCarthy {et~al.}(2003{\natexlab{a}})McCarthy, Babul, Holder, \&
  Balogh}]{McCarthy2003a}
McCarthy, I.~G., Babul, A., Holder, G.~P., \& Balogh, M.~L. 2003{\natexlab{a}},
  \apj, 591, 515

\bibitem[{McCarthy {et~al.}(2003{\natexlab{b}})McCarthy, Holder, Babul, \&
  Balogh}]{McCarthy2003b}
McCarthy, I.~G., Holder, G.~P., Babul, A., \& Balogh, M.~L. 2003{\natexlab{b}},
  \apj, 591, 526

\bibitem[{Molnar {et~al.}(2009)Molnar, Umetsu, Birkinshaw, Bryan, Haiman,
  Hearn, Ho, Huang, Koch, Liao, Lin, Liu, Nishioka, Wang, \& Wu}]{Molnar2008}
Molnar, S.~M. {et~al.} 2009, \apj, submitted

\bibitem[{Morandi {et~al.}(2007)Morandi, Ettori, \& Moscardini}]{Morandi2007}
Morandi, A., Ettori, S., \& Moscardini, L. 2007, \mnras, 379, 518, 0704.2678

\bibitem[{Motl {et~al.}(2005)Motl, Hallman, Burns, \& Norman}]{Motl2005}
Motl, P.~M., Hallman, E.~J., Burns, J.~O., \& Norman, M.~L. 2005, \apjl, 623,
  L63

\bibitem[{Mushotzky \& Scharf(1997)}]{Mushotzky1997}
Mushotzky, R.~F., \& Scharf, C.~A. 1997, \apj, 482, L13

\bibitem[{Myers {et~al.}(1997)Myers, Baker, Readhead, Leitch, \&
  Herbig}]{Myers1997}
Myers, S.~T., Baker, J.~E., Readhead, A. C.~S., Leitch, E.~M., \& Herbig, T.
  1997, \apj, 485, 1

\bibitem[{Nagai(2006)}]{Nagai2006}
Nagai, D. 2006, \apj, 650, 538

\bibitem[{Nagai {et~al.}(2007)Nagai, Kravtsov, \& Vikhlinin}]{Nagai2007}
Nagai, D., Kravtsov, A.~V., \& Vikhlinin, A. 2007, \apj, 668, 1,
  astro-ph/0703661

\bibitem[{Navarro {et~al.}(1997)Navarro, Frenk, \& White}]{Navarro1997}
Navarro, J.~F., Frenk, C.~S., \& White, S. D.~M. 1997, \apj, 490, 493

\bibitem[{Nishioka {et~al.}(2009)Nishioka, Wang, Wu, Ho, Huang, Koch, Liao,
  Lin, Liu, Molnar, Umetsu, Birkinshaw, Altamirano, Chang, Chang, Chang, Chen,
  Han, Huang, Hwang, Jiang, Kesteven, Kubo, Li, Martin-Cocher, Oshiro, Raffin,
  Wei, \& Wilson}]{Nishioka2008}
Nishioka, H. {et~al.} 2009, \apj, 694, 1637, 0811.1675

\bibitem[{Patel {et~al.}(2000)Patel, Joy, Carlstrom, Holder, Reese, Gomez,
  Hughes, Grego, \& Holzapfel}]{Patel2000}
Patel, S.~K. {et~al.} 2000, \apj, 541, 37

\bibitem[{Press {et~al.}(2002)Press, Teukolsky, Vetterling, \&
  Flannery}]{Press2002}
Press, W.~H., Teukolsky, S.~A., Vetterling, W.~T., \& Flannery, B.~P. 2002,
  Numerical Recipes in C++. The Art of Scientific Computing, 2nd edn.
  (Cambridge University Press)

\bibitem[{Reese {et~al.}(2002)Reese, Carlstrom, Joy, Mohr, Grego, \&
  Holzapfel}]{Reese2002}
Reese, E.~D., Carlstrom, J.~E., Joy, M., Mohr, J.~J., Grego, L., \& Holzapfel,
  W.~L. 2002, \apj, 581, 53

\bibitem[{Reiprich \& B\"{o}hringer(2002)}]{Reiprich2002}
Reiprich, T.~H., \& B\"{o}hringer, H. 2002, \apj, 567, 716

\bibitem[{Sanderson \& Ponman(2003)}]{Sanderson2003}
Sanderson, A. J.~R., \& Ponman, T.~J. 2003, \mnras, 345, 1241

\bibitem[{Sunyaev \& Zel'dovich(1970)}]{SZ1970}
Sunyaev, R.~A., \& Zel'dovich, Y.~B. 1970, Comments Astrophys. Space Phys., 2,
  66

\bibitem[{Udomprasert {et~al.}(2004)Udomprasert, Mason, Readhead, \&
  Pearson}]{Udomprasert2004}
Udomprasert, P.~S., Mason, B.~S., Readhead, A.~C.~S., \& Pearson, T.~J. 2004,
  \apj, 615, 63

\bibitem[{Umetsu {et~al.}(2009)Umetsu, Birkinshaw, Liu, Wu, Medezinski,
  Broadhurst, Lemze, Zitrin, Ho, Huang, Koch, Liao, Lin, Molnar, Nishioka,
  Wang, Altamirano, Chang, Chang, Chang, Chen, Han, Huang, Hwang, Jiang,
  Kesteven, Kubo, Li, Martin-Cocher, Oshiro, Raffin, Wei, \&
  Wilson}]{Umetsu2008}
Umetsu, K. {et~al.} 2009, \apj, 694, 1643, 0810.0969

\bibitem[{Vikhlinin(2006)}]{Vikhlinin2006a}
Vikhlinin, A. 2006, \apj, 640, 710

\bibitem[{Wu {et~al.}(2009)Wu, Ho, Huang, Koch, Liao, Lin, Liu, Molnar,
  Nishioka, Umetsu, Wang, Altamirano, Chang, Chang, Chang, Chen, Han, Huang,
  Hwang, Jiang, Kesteven, Kubo, Li, Martin-Cocher, Oshiro, Raffin, Wei, \&
  Wilson}]{Wu2008}
Wu, J.~H.~P. {et~al.} 2009, \apj, 694, 1619, 0810.1015

\end{thebibliography}

\end{document}